\documentclass[aps,twocolumn,floatfix,superscriptaddress,prl]{revtex4-1}
\usepackage{mathrsfs}
\usepackage{color}
\usepackage{graphicx}           
\usepackage{dcolumn}            
\usepackage{bm}                 
\usepackage{hyperref}           
\usepackage[latin1]{inputenc}   
\usepackage{pstricks}           
\usepackage{amsmath,amssymb}
\usepackage{version}
\usepackage{epstopdf}
\usepackage{lipsum}
\DeclareGraphicsExtensions{.png,.eps,.ps}
\usepackage{ulem}
\usepackage{float}
\usepackage{booktabs}
\usepackage{subfigure}
\usepackage{graphicx}
\usepackage{dcolumn}
\usepackage{bm}

\usepackage{float}

\definecolor{red}{rgb}{1,0,0}

\begin{document}

    \title{Magnetic-field effect in high-order above-threshold ionization}
    
    \author{Kang Lin}\email{lin@atom.uni-frankfurt.de}\address{Institut f\"ur Kernphysik, Goethe-Universit\"at Frankfurt am Main, Frankfurt am Main 60438, Germany}\address{State Key Laboratory of Precision Spectroscopy, East China Normal University, Shanghai 200241, China}
    \author{Simon Brennecke}\address{Institut f\"ur Theoretische Physik, Leibniz Universit\"at Hannover, Hannover 30167, Germany}
    \author{Hongcheng Ni}\address{State Key Laboratory of Precision Spectroscopy, East China Normal University, Shanghai 200241, China}\address{Institute for Theoretical Physics, Vienna University of Technology, Vienna 1040, Austria}
    \author{Xiang Chen}\address{Key Laboratory for Laser Plasmas (Ministry of Education) and School of Physics and Astronomy, Collaborative innovation center for IFSA (CICIFSA), Shanghai Jiao Tong University,  Shanghai 200240, China}
    \author{Alexander Hartung}\address{Institut f\"ur Kernphysik, Goethe-Universit\"at Frankfurt am Main, Frankfurt am Main 60438, Germany}
    \author{Daniel Trabert}\address{Institut f\"ur Kernphysik, Goethe-Universit\"at Frankfurt am Main, Frankfurt am Main 60438, Germany}    
    \author{Kilian Fehre}\address{Institut f\"ur Kernphysik, Goethe-Universit\"at Frankfurt am Main, Frankfurt am Main 60438, Germany} 
    \author{Jonas Rist}\address{Institut f\"ur Kernphysik, Goethe-Universit\"at Frankfurt am Main, Frankfurt am Main 60438, Germany} 
    \author{Xiao-Min Tong}\address{Center for Computational Sciences, University of Tsukuba, Tsukuba, Ibaraki 305-8573, Japan}     
    \author{Joachim Burgd\"orfer}\address{Institute for Theoretical Physics, Vienna University of Technology, Vienna 1040, Austria}        
    \author{Lothar. Ph. H. Schmidt}\address{Institut f\"ur Kernphysik, Goethe-Universit\"at Frankfurt am Main, Frankfurt am Main 60438, Germany}         
    \author{Markus S. Sch\"offler}\address{Institut f\"ur Kernphysik, Goethe-Universit\"at Frankfurt am Main, Frankfurt am Main 60438, Germany}  
    \author{Till Jahnke}\address{Institut f\"ur Kernphysik, Goethe-Universit\"at Frankfurt am Main, Frankfurt am Main 60438, Germany}
    \author{Maksim Kunitski}\address{Institut f\"ur Kernphysik, Goethe-Universit\"at Frankfurt am Main, Frankfurt am Main 60438, Germany}    
    \author{Feng He}\address{Key Laboratory for Laser Plasmas (Ministry of Education) and School of Physics and Astronomy, Collaborative innovation center for IFSA (CICIFSA), Shanghai Jiao Tong University,  Shanghai 200240, China}       
    \author{Manfred Lein}\address{Institut f\"ur Theoretische Physik, Leibniz Universit\"at Hannover, Hannover 30167, Germany}     
    \author{Sebastian Eckart}\address{Institut f\"ur Kernphysik, Goethe-Universit\"at Frankfurt am Main, Frankfurt am Main 60438, Germany}  
    \author{Reinhard D\"orner}\email{doerner@atom.uni-frankfurt.de} \address{Institut f\"ur Kernphysik, Goethe-Universit\"at Frankfurt am Main, Frankfurt am Main 60438, Germany}

    \date{\today}

    \begin{abstract}
        {
We experimentally and theoretically investigate the influence of the magnetic component of an electromagnetic field on high-order above-threshold ionization of xenon atoms driven by ultrashort femtosecond laser pulses. The nondipole shift of the electron momentum distribution along the light-propagation direction for high energy electrons beyond the classical cutoff is found to be vastly different from that below the cutoff. A V-shape structure in the momentum dependence of the nondipole shift above the cutoff is identified for the first time. With the help of classical and quantum-orbit analysis, we show that large-angle rescattering of the electrons strongly alters the partitioning of the photon momentum between electron and ion. The sensitivity of the observed nondipole shift to the electronic structure of the target atom is confirmed by three-dimensional time-dependent Schr\"odinger equation simulations for different model potentials.}

    \end{abstract}

    \maketitle

Conservation of energy and momentum are among the most fundamental laws of physics. Still it is often a non-trivial question how they manifest for a given microscopic quantum process. An intriguing scenario to inspect this circumstance is the light-driven ionization of atoms and molecules where the absorption of energy $E_\gamma$ from the electromagnetic field traveling in $x$-direction goes along with the transfer of a linear momentum of $\Delta p_x=E_\gamma/c$~\cite{Smeenk}, where $c$ is the speed of light (atomic units are used unless stated otherwise). This momentum is absorbed by the center of mass of the photo fragments (electrons and ions) in each ionization event. The photon momentum is neglected in the frequently used dipole approximation of light-matter interaction~\cite{Maurer_2021}. Hence, within this approximation the expectation values of the electron momentum $\langle p_{x,e} \rangle$ and ion momentum $\langle p_{x,ion} \rangle$ along the light-propagation direction are zero for the ionization of atoms. Momentum conservation, however, dictates for each kinetic energy $E_e$ and ionization potential $I_p$ (neglecting the kinetic energy of the ion):

\begin{equation}\label{eq1}
\left \langle p_{x,e} \right \rangle+\left \langle p_{x,ion} \right \rangle=\frac{E_\gamma}{c}=\frac{E_{e}}{c}+\frac{I_{p}}{c}.
\end{equation}

This kinematic conservation law from Eq. (\ref{eq1}) leaves open how the photon momentum is shared between electron and ion. This depends on the dynamics of the ionization process. For example, at high photon energies one finds $\langle p_{x,e} \rangle = 8/5\times E_e/c$ and $\langle p_{x,ion}\rangle =-3/5\times E_e/c+I_p/c$ for single-photon ionization in the perturbative regime~\cite{Sommerfeld,Chelkowski,Grundmann}. For recollision-free strong-field ionization, e.g., by circularly polarized light, the simplest estimate is $\langle p_{x,e}\rangle \approx E_e/c+1/3\times I_p/c$~\cite{Klaiber,Pei-Lun,Hartung2019}. Mechanistically, this can be understood as the classical Lorentz force induced by the magnetic field acting on a free electron which has been accelerated by the electric field in the polarization plane ($yz$-plane) resulting in the $E_e/c$ term. The additional offset of $I_p/(3c)$ is caused by the action of the magnetic field on the electron while it is set free (i.e. by tunneling).

In contrast to the recollision-free ionization, in case of employing linearly polarized laser fields the electrons may be driven back to the parent ion and undergo scatterings with the ionic core. This leads to a fraction of low-energy electrons which are Coulomb-focused and form a backward-shifted narrow cusp in the momentum distribution ~\cite{Ludwig2014,Chelkowski2015,Haram2019,Maurer2018}. Additionally, photoelectron holography~\cite{Joachim2006,Huismans2011}, i.e., the interference between direct and forward scattered electrons, dominates the distributions at intermediate energies below the classical cutoff of $2U_p$, where $U_p=E_0^2/(4\omega^2)$ is the ponderomotive potential of the driving laser field with electric field amplitude $E_0$ and frequency $\omega$. In this region, the electron momentum $\langle p_{x,e} \rangle$ is approximately given by $\langle p_{x,e}\rangle \approx E_{e}/c$~\cite{Hartung2019,Brennecke2019PRA,Willenberg2019}. Classically, energies higher than $2U_p$ can only be reached by rescattering electrons~\cite{Paulus1994PRL,Paulus1994JPB}, i.e., electrons that are elastically scattered into large angles followed by further acceleration in the laser field. In this high-order above-threshold ionization (HATI) process,  electrons can reach energies of up to $10 U_p$~\cite{Paulus1994PRL,Paulus1994JPB}. These high-energy electrons are at the heart of many well-known phenomena in strong-field physics, including high-harmonic generation~\cite{McPherson1987,Ferray_1988,Paul2001} and laser-induced electron diffraction~\cite{Lein2002,Morishita2008,Okunishi2008,Ray2008}, which both are, for example, important for atomic and molecular imaging~\cite{Meckel,blaga2012,pullen2015}. The inherent requirements of high laser intensity and long wavelength for accelerating high-energy rescattering electrons naturally places these processes outside the dipole oasis~\cite{Reiss2014} and, hence, challenges the applicability of the dipole approximation that was widely adopted in previous studies. Recently, the understanding of nondipole effects in strong-field ionization has been deepened by analyzing effects such as nonsequential double ionization~\cite{Emmanouilidou2017,Sun_2020,ChenXiang}, the electric-field inhomogeneity~\cite{Hartung2021}, nonadiabaticity \cite{Hongcheng2020} and weak Coulomb effects~\cite{willenberg2019sub,he2021nondipole}.

Nevertheless, it is still an open question how the photon momentum is shared between the electron and ion in HATI. Brennecke \textit{et al}.~\cite{Brennecke2018} were the first to predict the nondipole shift for HATI of helium atoms finding that $\langle p_{x,e} \rangle$ levels off and remains constant beyond the $2U_p$ cutoff. This prediction has remained experimentally untested so far. In this Letter, we present the first experimental results on the photon-momentum transfer for the ionization of xenon atoms by linearly polarized light reaching energies beyond the classical $2U_p$ cutoff. Even though we find indeed a breakdown of the linear scaling of $\langle p_{x,e} \rangle$ with $E_e/c$ as soon as the electron energy exceeds $2U_p$, our results for the nondipole shift in the HATI regime are at variance with the prediction for helium. We find that the amount of transferred photon momentum strongly depends on the details of the scattering process. Hence, the momentum transfer is highly target sensitive. In case of xenon atoms, a V-shape structure in the momentum dependence of the nondipole shift is identified for the first time.

The observation of the linear-photon-momentum transfer in strong-field ionization is challenging due to its extremely small magnitude, which is only $4\times10^{-4}$~a.u. for a single photon of typical table-top Ti:Sapphire laser systems that work at central wavelengths of about 800~nm. We meet that challenge by adopting the same strategy of counter-propagating laser pulses as described in Refs. ~\cite{Hartung2019,Hartung2021}. We only give a brief outline here. The output (25~fs, 800~nm, 10~kHz) of a Ti:Sapphire laser system (Coherent Legend Elite Duo) was split into two pathways using a dielectric beam splitter, after which the intensity and polarization of each beam can be adjusted independently. The two beams were guided into the vacuum chamber of a cold target recoil ion momentum spectroscopy (COLTRIMS) system ~\cite{DORNER200095} from two opposite directions and focused onto the same spot inside a supersonic gas jet of xenon atoms by two independent lenses ($f=25$~cm). The peak intensity in the reaction volume was estimated to be $7\times 10^{13}$~W/cm$^2$ ($U_p=4.2$~eV). A static electric field of 29.8~V/cm was applied to guide the created electrons and ions to two time- and position-sensitive detectors at the opposite ends of the spectrometer. The three-dimensional momenta of the electrons and ions were retrieved in coincidence from the times-of-flight and positions-of-impact. Owing to the mirror symmetry of the experiment with respect to the light polarization axis ($z$-axis), the data were symmetrized in that dimension. 

\begin{figure}
    \includegraphics[width=1.0\columnwidth]{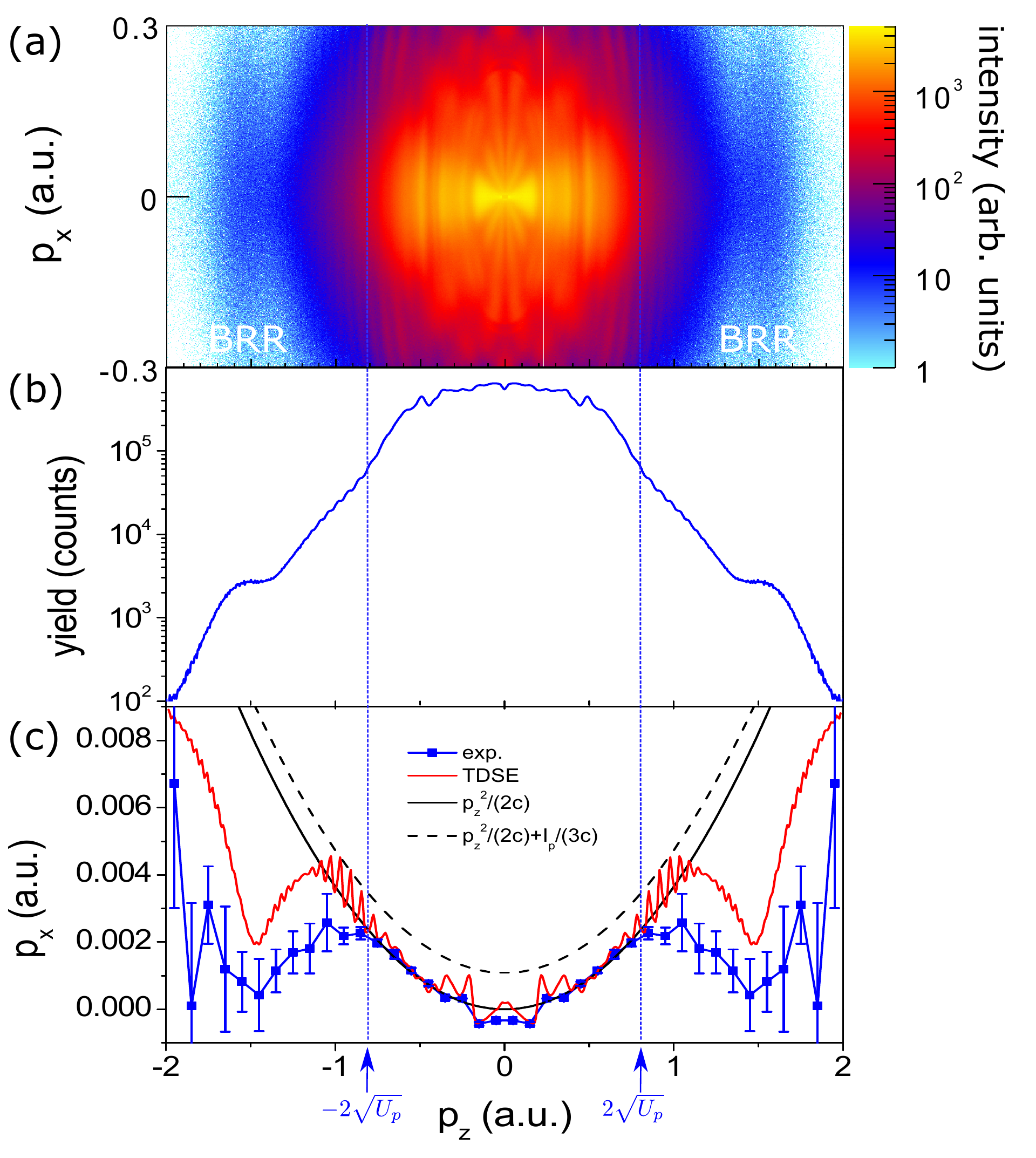}
    \caption{Nondipole effects in strong-field ionization of xenon atoms (linear polarization, $7\times10^{13}$~W/cm$^2$, 800~nm, 25~fs). (a) Experimentally measured two-dimensional momentum distribution in the plane formed by polarization axis ($z$-axis) and light-propagation direction ($x$-axis). (b) One-dimensional momentum distribution along the polarization axis, where a plateau structure is typical for high-order above-threshold ionization (HATI). (c) Peak electron momentum in the light-propagation direction $p_x$ as a function of the momentum along the polarization axis $p_z$. The error bars show statistical errors. The TDSE result calculated for the Tong-Lin potential~\cite{Zhang2014,Tong2017} is averaged over the carrier-envelope phase as well as the focal intensity distribution assuming a peak intensity of $7 \times 10^{13}$~W/cm$^2$.}
    \label{exp}
\end{figure}

Figure \ref{exp}(a) displays the measured two-dimensional photoelectron momentum distribution along the polarization axis ($z$-axis) and light-propagation direction ($x$-axis). Here, three regions can be clearly identified: At low energies with $|p_z|\lessapprox 0.2$~a.u., the electrons experience strong Coulomb focusing induced by the parent ion and form a narrow cusp. For the intermediate region with $0.2\; \text{a.u.}\lessapprox|p_z| \lessapprox2\sqrt{U_p}$, the distribution is dominated by photoelectron holography and also ATI rings are visible. For higher energies with $|p_z|\gtrapprox2\sqrt{U_p}$, the distribution is dominated by HATI. Close to the classical $10U_p$ cutoff for HATI around $|p_z|\approx1.5$~a.u., a ridge structure can be identified, which is termed backward rescattering ridge (BRR)~\cite{Morishita2008}. This can be seen more clearly as a plateau-like structure in the projected momentum distribution along the polarization axis shown in Fig. \ref{exp}(b) that is calculated by integration over the light-propagation direction. To quantify the photon-momentum transfer, we determine the peak position of the momentum distribution along the light-propagation direction for each $p_z$ by fitting a Gaussian function to the central region with $|p_x|<0.2$~a.u. The result is shown in Fig. \ref{exp}(c). For intermediate energies below $2U_p$, the peak position coincides well with the prediction $p_z^2/(2c)$ and is in agreement with an earlier experiment~\cite{Hartung2019}. For low-energy electrons, the peak in the momentum distribution shifts towards negative values ~\cite{Ludwig2014,Chelkowski2015,Haram2019,Maurer2018}. However, for electrons with $|p_z|\gtrapprox2\sqrt{U_p}$, the peak position deviates drastically from the expected parabolic shape. In contrast, a pronounced V-shape structure appears in the region dominated by HATI. 

\begin{figure}
    \includegraphics[width=1.0\columnwidth]{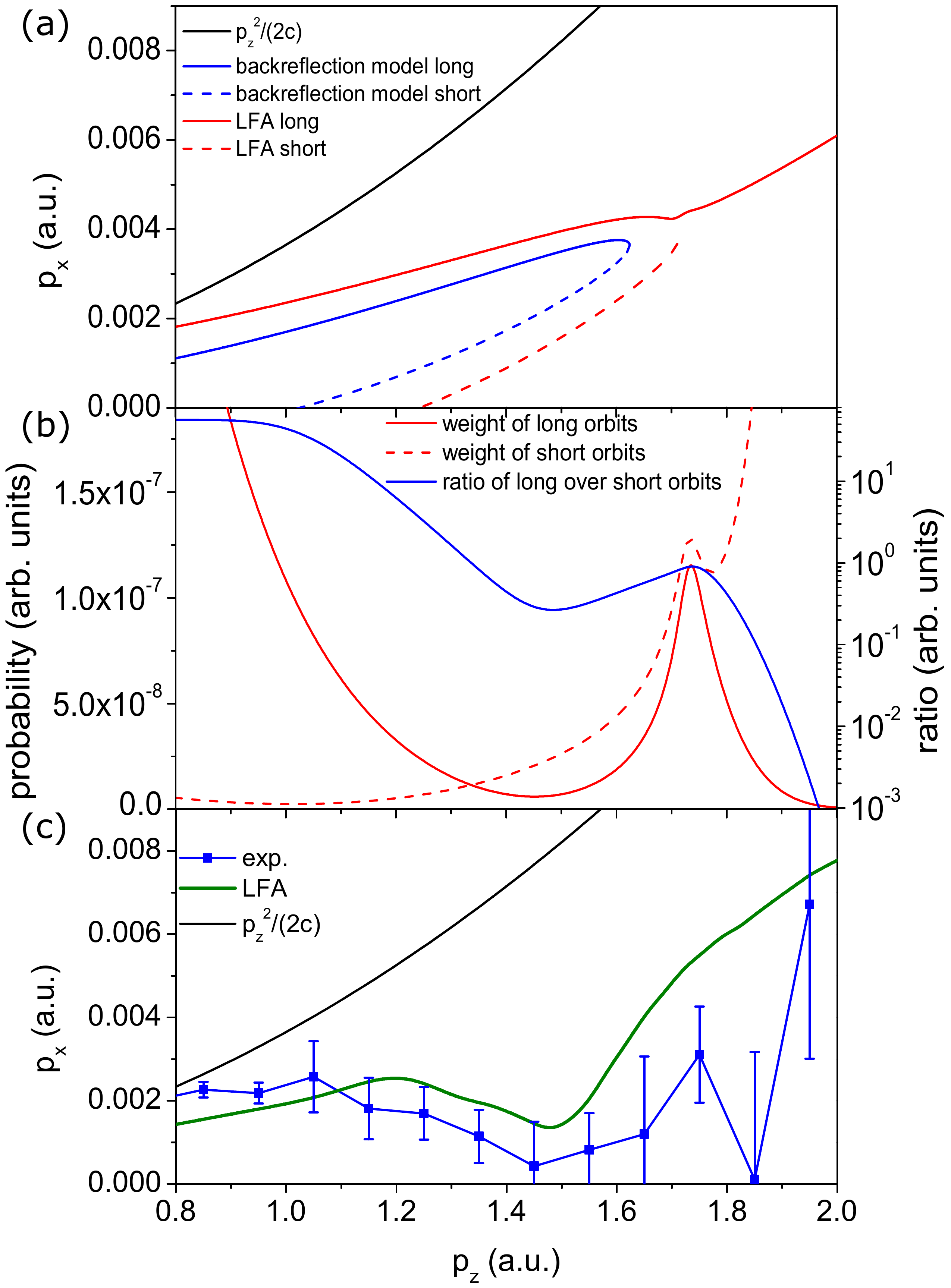}
    \caption{Results of the classical and quantum-orbit analysis. (a) Momentum-shift estimate from the classical backreflection model and low-frequency approximation (LFA) for electrons scattered by $180^\circ$. (b) Probabilities of quantum orbits with final momentum $p_x=0$ in LFA. In (a) and (b), the results for the long and short trajectories are shown as solid and dashed lines for an intensity of $6\times 10^{13}$ W/cm$^2$. (c) Focal-volume-averaged LFA result calculated with the differential cross section for the Tong-Lin potential including both long and short orbits (green line) and the experimental result (blue line).}
    \label{LFA}
\end{figure}

To reveal the underlying physics of the V-shape structure, we start with a simple classical backreflection model that is based on a nondipole generalization \cite{Brennecke2018} of the three-step model~\cite{Corkum1994,Paulus1994JPB}. Here, the ionization step launches an electron at time $t_i$ with vanishing initial velocity. Afterwards, the freed electron is driven by the electromagnetic laser field and the ionic potential is neglected. Some trajectories are finally driven back close to the ionic core at a recollision time $t_r$ and may scatter elastically in a third step. However, the magnetic-field induced Lorentz force rotates the velocity vector giving rise to a component of the incident velocity along the light-propagation direction  $v_{{\rm in},x}=v_{{\rm in},z}^2/(2c)$ in addition to the velocity component $v_{in,z}$ along the polarization axis, which breaks the forward-backward symmetry. The differential electron scattering cross section of Xe$^+$ ion has a local maximum at a scattering angle of $180^\circ$~\cite{Morishita2008} (backscattering) which induces the maximum in $p_x$-direction near the polarization axis in the momentum distribution. We therefore estimate the nondipole shift by tracking only the exactly backward scattered electrons by reversing their velocity vector during the scattering process. At this instant, twice the electron momentum is transferred to the parent ion. Although the electrons are afterwards further accelerated in the laser field, the scattering process causes a considerable reduction of the final electron momentum component along the light-propagation direction. The result of the backreflection model is shown in Fig. \ref{LFA}(a). Here, two trajectories born per optical half cycle contribute to the same final momentum component in polarization direction $p_z$, which are referred to as the ``long'' and ``short'' trajectories, respectively~\cite{Corkum1994}. Strikingly, the nondipole shifts of both trajectories are significantly smaller than $p_z^2/(2c)$, which agrees qualitatively with the experimental observation. However, the backreflection model cannot fully explain all details such as the V-shape structure.

For a more realistic modeling, we implement a low-frequency approximation (LFA)~\cite{milosevic2014} based on the quantum-orbit analysis within the strong-field approximation but beyond the electric dipole approximation~\cite{Brennecke2018PRA}. The basic ingredient of the LFA is complex-valued electron trajectories, and it also accurately incorporates the field-free quantum-mechanical differential cross section for electron rescattering. The probability amplitude for a single quantum orbit with final momentum $\textbf{p}$ is given by \begin{equation}
M(\textbf{p})=P(\textbf{p})\,T (\textbf{v}_{\rm in}(\textbf{p}), \textbf{v}_{\rm out}(\textbf{p})).
\end{equation}
In the HATI region, the shape of the photoelectron momentum distribution is mostly determined by the T-matrix element that is related to the differential cross section $\sigma\propto |T|^2$. Here, the arguments $\textbf{v}_{\rm in}(\textbf{p})$ and $\textbf{v}_{\rm out}(\textbf{p})$ are electron velocities shortly before and after scattering~\cite{Brennecke2018PRA}, respectively. All target-independent factors are collected in the prefactor~$P$. For trajectories scattered by $180^\circ$, the short and long quantum orbits show a nondipole shift similar to the classical backreflection model, see Fig. \ref{LFA}(a). However, due to the nonadiabatic initial velocity we find that the difference of the nondipole shift between short and long quantum orbits is larger. Since both quantum orbits contribute to the final momentum distribution, the nondipole shift is strongly influenced by the relative weight between the two trajectories, as shown in Fig. \ref{LFA}(b). The relative weight is mainly determined by the energy-dependent differential cross section. The long quantum orbits dominate the momentum distribution for $p_z \lessapprox 1.3$~a.u., while the short ones become more prominent for higher $p_z$ up to the cutoff. Around the cutoff at $p_z\approx 1.7$~a.u. both orbits become almost indistinguishable and for even higher $p_z$ beyond that cutoff only the long trajectory is physically meaningful. In the calculation, these constraints are ensured by using the uniform approximation~\cite{Becker2002}. The transition from predominately long to short and back to long trajectories leads to a minimum, as indicated in the ratio curve in Fig. \ref{LFA}(b), which finally results in the V-shape structure in the nondipole shift $p_x$ as a function of $p_z$. The focal-volume averaged result of the LFA simulation is shown in Fig. \ref{LFA}(c), which quantitatively reproduces the experimentally observed V-shape structure.

\begin{figure}
  	\includegraphics[width=1.0\columnwidth]{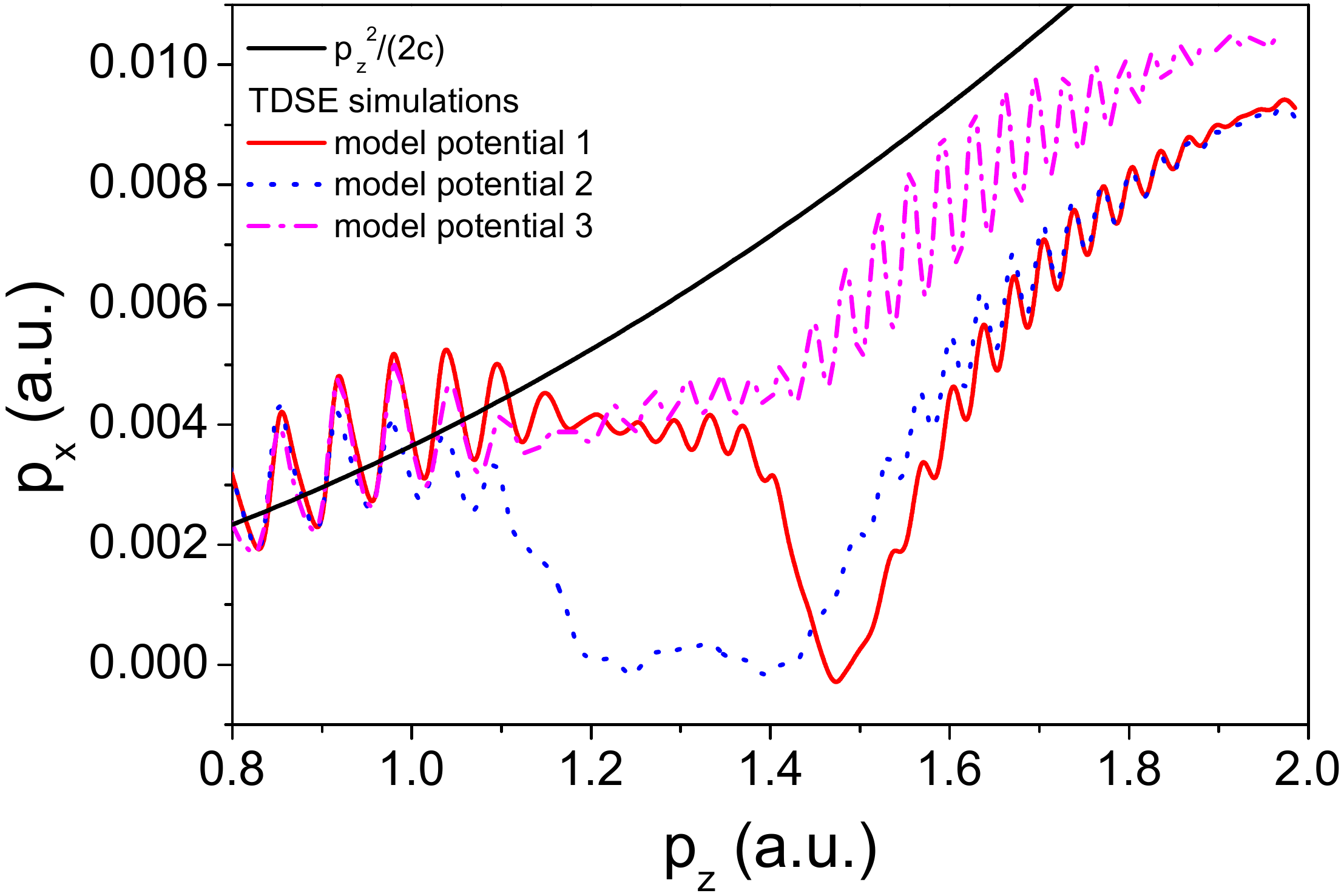}
  	\caption{Model potential dependence of the nondipole shift. Presented are TDSE calculations for three different single-active-electron potentials for xenon: Tong-Lin potential, see \cite{Zhang2014}, (model potential 1), Green-Sellin-Zachor (GSZ) potential, see \cite{Zhang2014,GSZ}, (model potential 2) and the effective potential $V (r) = -(Z + a_1e^{-a_2r})/r$, where $Z = 1, a_1 = 1.985, a_2 = 0.5$ (model potential 3). Here, the results are averaged over the carrier-envelope phase, but only a single intensity of $6\times10^{13}$~W/cm$^2$ is used.}
  	\label{TDSE}
\end{figure}

In order to undergo backscattering, the electron has to return close to the parent ion, which renders the HATI process much more sensitive to the ionic potential than recollision-free ionization. As discussed above, the LFA simulation indicates that the nondipole shift strongly depends on the exact differential cross section, i.e., the details of the interaction between the active electron and the ionic core. This explains the striking difference between the present result and the earlier prediction for a helium model atom~\cite{Brennecke2018}, where a flat plateau in the nondipole shift was predicted. In order to test for the dependence on the ionic potential, we have performed more elaborate calculations by numerically solving the three-dimensional time-dependent Schr\"odinger equation (TDSE) without invoking the dipole approximation. To this end, the generalized pseudospectral method~\cite{TONG1997,Tong2017,Gao2019} has been extended to the nondipole regime by expanding the nondipole contribution in a Taylor series~\cite{Hongcheng2020}. Figure \ref{TDSE} shows the TDSE results for three different single-active-electron potentials. All the potentials produce the correct ionization potential and have the same asymptotic form of $-1/r$, but differ in the inner region. The predicted nondipole shifts are quite different in the HATI region, which implies that the electron-core interaction in the inner-region plays an important role for the shifts. In comparison with the experiment, model potential 1, i.e. the Tong-Lin potential~\cite{Tong_2005,Tong1997PRA}, gives the most accurate results. In contrast, both model potentials 2 and 3 fail to describe the interaction of the active electron with the core at short distances. Our comparison shows that the nondipole shifts for backscattering electrons are very sensitive to the details of the scattering process. This holds the promise of decoding the detailed interaction of the active electron with the ionic core at short distance. 

In conclusion, we show that the partitioning of the linear photon momentum between electron and nucleus in strong-field ionization is drastically altered by the rescattering process. For electrons with energies above the classical 2$U_p$ cutoff, the photon momentum sharing differs drastically from that of energy below 2$U_p$. For the xenon target investigated here, the momentum distribution in the relevant region is dominated by backscattering electrons. The recollision of the electron on the parent ion reverses the electron momentum vector and thus, twice the electron momentum (before scattering) is transferred to the ion. This causes a considerable reduction of the final electron momentum in light-propagation direction. As a result, our work provides the first observation of a target-sensitive nondipole effect in strong-field ionization. The sensitivity of the nondipole shift to the model potential of the parent ion in turn provides us with a promising tool for atomic or molecular imaging.

\begin{acknowledgments}
The experimental work was supported by the DFG (German Research Foundation).
K. L. acknowledges support by the Alexander von Humboldt Foundation and thanks Xiaoqing Hu for helpful discussion. S. E. acknowledges funding of the DFG through Priority Programme SPP 1840 QUTIF. A. H. and K. F. acknowledge support by the German Academic Scholarship Foundation. X. M. T. was supported  by Multidisciplinary Cooperative Research Program in CCS, University of Tsukuba. F. H. acknowledges the support by the National Science Foundation of China (No. 11925405). H. N. acknowledges the support by Project No. 11904103 of the National Natural Science Foundation of China (NSFC), Project No. M2692 of the Austrian Science Fund (FWF), and Projects No. 21ZR1420100 and 19JC1412200 of the Science and Technology Commission of Shanghai Municipality. Numerical simulations were in part performed on the ECNU Multifunctional Platform for Innovation (001). 
\end{acknowledgments}

\bibliographystyle{apsrev4-1}
\bibliography{ref}

\end{document}